%% file: giammanco_exp_summary.tex
\documentclass[12pt]{article}
\usepackage{graphicx}
\usepackage{hyperref}
\usepackage{xspace}

\newcommand\pubnumber{CP3-17-53}
\newcommand\pubdate{\today}

\newcommand{\pb}{\ensuremath{{\rm pb}^{-1}}\xspace}
\newcommand{\fb}{\ensuremath{{\rm fb}^{-1}}\xspace}

\newcommand{\pp}{\ensuremath{pp}\xspace}
\newcommand{\ppbar}{\ensuremath{p\bar{p}}\xspace}
\newcommand{\ttbar}{\ensuremath{t\bar{t}}\xspace}
\newcommand{\ttH}{\ensuremath{\ttbar H}\xspace}

\def\institute{Centre for Cosmology, Particle Physics and Phenomenology (CP3)\\
Universit\'e catholique de Louvain\\
Chemin du Cyclotron 2, B-1348 Louvain-la-Neuve, Belgium}

\def\Title#1{\begin{center} {\Large #1 } \end{center}}
\def\Author#1{\begin{center}{ \sc #1} \end{center}}
\def\Address#1{\begin{center}{ \it #1} \end{center}}

\newcommand\pubblock{\rightline{\begin{tabular}{l} \pubnumber\\
         \pubdate  \end{tabular}}}
\newenvironment{Abstract}{\begin{quotation}  }{\end{quotation}}
\newenvironment{Presented}{\begin{quotation} \begin{center} 
             PRESENTED AT\end{center}\bigskip 
      \begin{center}\begin{large}}{\end{large}\end{center} \end{quotation}}
\def\Acknowledgements{\bigskip  \bigskip \begin{center} \begin{large}
             \bf ACKNOWLEDGEMENTS \end{large}\end{center}}

\input econfmacros.tex

\begin{document}
\begin{titlepage}
\pubblock

\vfill
\Title{TOP2017: Experimental Summary}
\vfill
\Author{ Andrea Giammanco}
\Address{\institute}
\vfill
\begin{Abstract}
At the time of the 10th International Workshop on Top Quark Physics (TOP2017), top quark physics is a very dynamic research area.
Thanks to the unprecedentedly fast accumulation of high-energy data at the LHC during the ongoing Run~2,  statistical starvation is a matter of the past for most of the traditional top-quark analyses, that are now experiencing the luxury of having to worry about how to punch through the ``Systematics Wall'' and maximize the utility of their data. 
New processes involving top quarks are being studied for the first time, and the good old pair-production processes are being explored in unusual settings, such as collisions involving heavy ions, or ``reference data'' collected by the LHC at relatively low centre-of-mass energy.  
The TOP2017 conference featured 37 talks delivered by experimental physicists, including seven in the ``Young Scientists Forum'' where young colleagues were given the opportunity to elaborate more deeply than usual on their own work. 
As it is impossible to do justice to all the experimental results presented at this conference while staying within a reasonable length, this document contains a very biased selection, mostly based on the personal taste of the author.
\end{Abstract}
\vfill
\begin{Presented}
$10^{th}$ International Workshop on Top Quark Physics\\
Braga, Portugal,  September 17--22, 2017
\end{Presented}
\vfill
\end{titlepage}
\def\thefootnote{\fnsymbol{footnote}}
\setcounter{footnote}{0}

\section{Introduction: the Age of Plenty}

Experimental top-quark physics was born during the Run~1 of the Tevatron collider\footnote{Although searches for this quark at previous accelerators had been already attempted, including an aborted discovery in 1984~\cite{rubbia}.}, lasted from 1992 to 1996, which delivered 0.12~\fb of data at a centre-of-mass (CM) energy of 1.8~TeV. That period culminated with the top-quark discovery~\cite{top-discovery} by the CDF and D0 collaborations. Run~2 at the Tevatron lasted from 2002 to 2011, delivering 10~\fb of data at a CM energy of 1.96~TeV.
The first three TOP conferences, in 2006, 2008 and 2010, saw the Tevatron alone in the experimental top-quark arena, making the top-quark studies enter the age of precision, while the members of the multi-purpose ATLAS~\cite{atlas} and CMS~\cite{cms} experiments were increasingly busy with the simulation-based preparation for the upcoming LHC data. Most top-quark measurements at Tevatron, at that time, were limited by the size of the selected dataset. With a lot of ingenuity, the Tevatron pioneers laid the ground in those years for most of today's top quark physics, including the early studies of single top-quark production~\cite{st-discovery}. 
Since TOP2011, four experimental collaborations~\footnote{But five collaborations, overall, presented top-quark measurements, including the LHCb collaboration with their measurement of $\ttbar+tW$ cross section in a fiducial volume complementary to the ATLAS and CMS acceptances~\cite{lhcb-top}.}, using data from two colliders, routinely present their results at this conference (annual since then). Each of the multi-purpose experiments at the LHC integrated a luminosity of about 5~\fb of good quality data at 7~TeV in 2010-2011, followed by about 20~\fb at 8~TeV in 2012~\footnote{I was very excited to prepare a talk at TOP2011, where I was invited to elaborate on the perspectives for top-quark physics if the LHC achieved its goal for 2012: accumulating around 10~\fb at 7~TeV. That speculative future dataset sounded fantastic at the time of the talk, but was already a matter of sniggering by the time I had to write the proceedings~\cite{giammanco-top2011}.},  and more than 80~\fb at 13 TeV by the time of writing this document (end of 2017). A recurring theme through this document is how the top-quark community is adjusting to this unprecedented Age of Plenty, where measurements that were statistics-limited until yesteryear are now hitting the ``Systematics Wall''.

These proceedings report the author's personal view of the interesting trends observed at the TOP2017 conference, and are organized as follows: 
Section~\ref{sec:production} is devoted to measurements of the abundances of top-quark production, by various mechanisms and in various multiplicity (pairs, single, and even four at the time), with or without associated particles, reported as inclusive or differential cross sections, or ratios. 
Section~\ref{sec:two-numbers} starts from taking seriously a question by some unnamed colleague from another field, ``Isn't the top quark just two numbers?'', as an excuse to delve into why those two numbers are so important after all. 
Section~\ref{sec:asymmetries} is entirely devoted to the measurement of angular distributions and asymmetries in top-quark events, as a possible pathway towards new Physics.


\section{Measuring the cornucopia}
\label{sec:production}

The general feeling at the time of TOP2017 is that most of the physics that can be done with top quark pairs and with the most abundant of the single top-quark channels has already hit what some physicists call the ``Systematics Wall'', i.e. the state where accumulation of more data (``reloading the analysis'') has very little direct benefit on the precision of a measurement or the sensitivity of a search. 
This is never a bad thing, as it forces people to be creative, and imposes more rigorous standards: when $stat \ll syst$, the practice of assigning conservative variations to some systematics goes from being pragmatical to being suicidal.

What is to be done? To answer the burning questions of our field~\cite{lenin}, and test the Standard Model (SM) in novel ways, the analysts squeeze additional knowledge from the same data in several different ways. For example:
\begin{itemize}
\item Take ratios;
\item Go differential;
\item Stop and think.
\end{itemize}


\subsection{Top-quark pairs: cross sections and their usage}

\begin{figure}[!h!tbp]
  \center
  \includegraphics[width=0.8\textwidth]{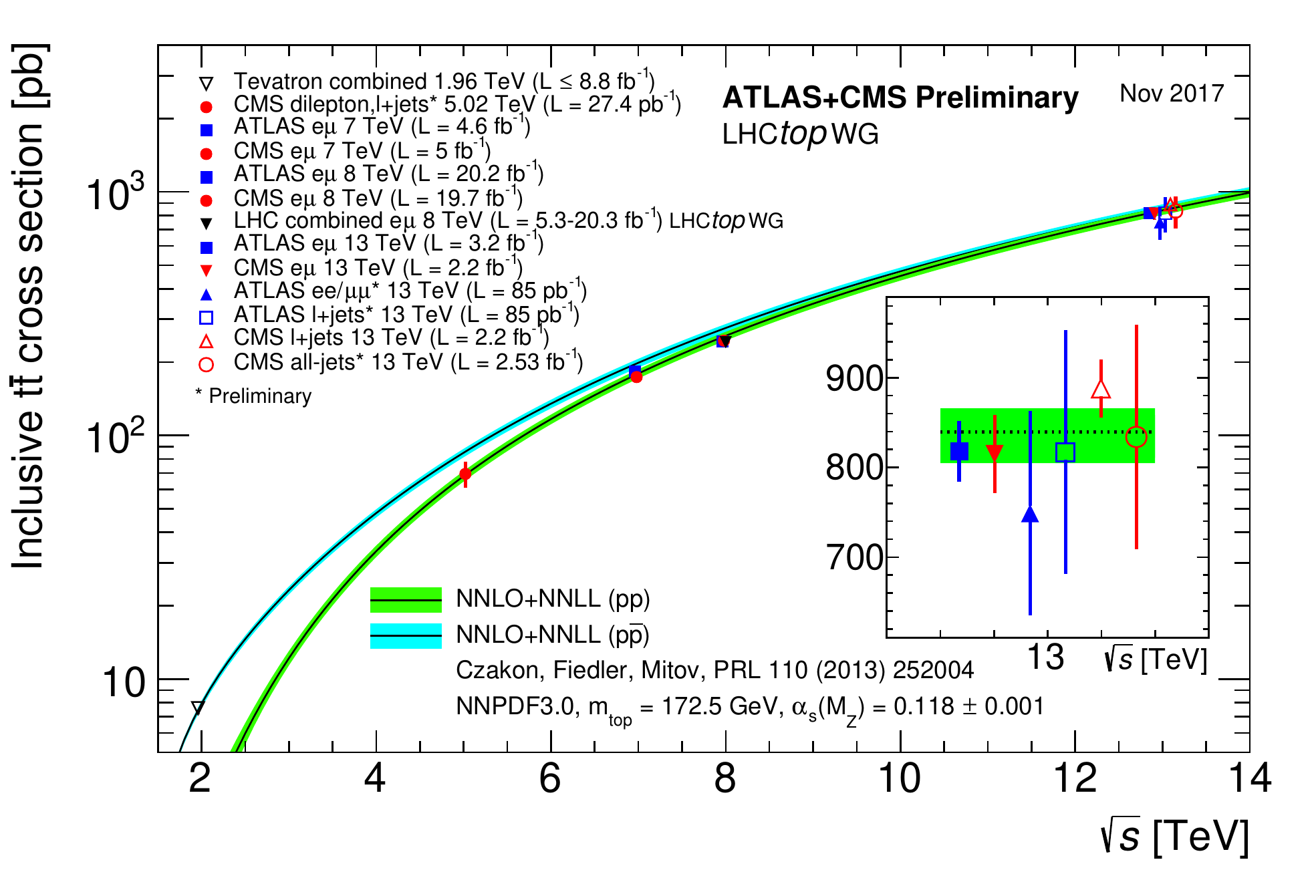}
  \caption{%
   Selected measurements of inclusive \ttbar cross sections in \pp and \ppbar collisions.  
From the LHC Top Working Group~\cite{lhctopwg}.
  }
  \label{fig:ttxsec}
\end{figure}

Figure~\ref{fig:ttxsec} shows the most precise measurements of inclusive \ttbar cross sections in \pp and \ppbar collisions~\cite{proc-ttbar-incl}. 
The latest addition, released at TOP2017, is the ATLAS measurement at 13~TeV in the single-leptonic channel~\cite{atlas-inclusive}. 
With the only exception of 5.02~TeV, all LHC measurements have hit the Systematics Wall. 

Most notably, luminosity uncertainty is limited to about 2\% for both multipurpose experiments and  features as the largest single uncertainty all across the board. 
In fact, asymptotically it is realistic to guess that \ttbar fiducial cross sections may come to be used as monitors of the stability of the luminosity calibration: the entire statistics collected in 2015 corresponds to about ten good LHC fills in 2017 (roughly one good week of data taking). With even higher intensities to be delivered in future LHC runs, traditional offline monitoring methods based on $Z$ bosons will become harder and harder to sustain (suitable triggers will be more and more difficult to keep unprescaled), and \ttbar production could be a good replacement. 
For example, one could break the dataset in chunks of about 2 or 3~\fb, i.e. the same size of the samples collected in 2015, which would yield a statistical uncertainty of about 1\% in the clean dileptonic final state~\cite{atlas_cms_dilepton_2015}; the \ttbar modeling uncertainties that dominate those cross section measurements are independent of instantaneous luminosity and would therefore have no impact on the relative calibration. On the other hand, systematics from detector effects and backgrounds (which affect the 2015 data analyses by an amount comparable to statistical uncertainty) would not cancel out exactly, as they have an indirect dependence on pileup, hence on luminosity itself, introducing some circularity that has to be carefully deconvoluted (as we already have to do for the standard method based on $Z$ bosons). In conclusion, \ttbar rates may be the {\it relative luminometer} of the future, for LHC and possible future high-intensity hadron colliders.

\begin{figure}[!h!tbp]
  \includegraphics[width=0.5\textwidth]{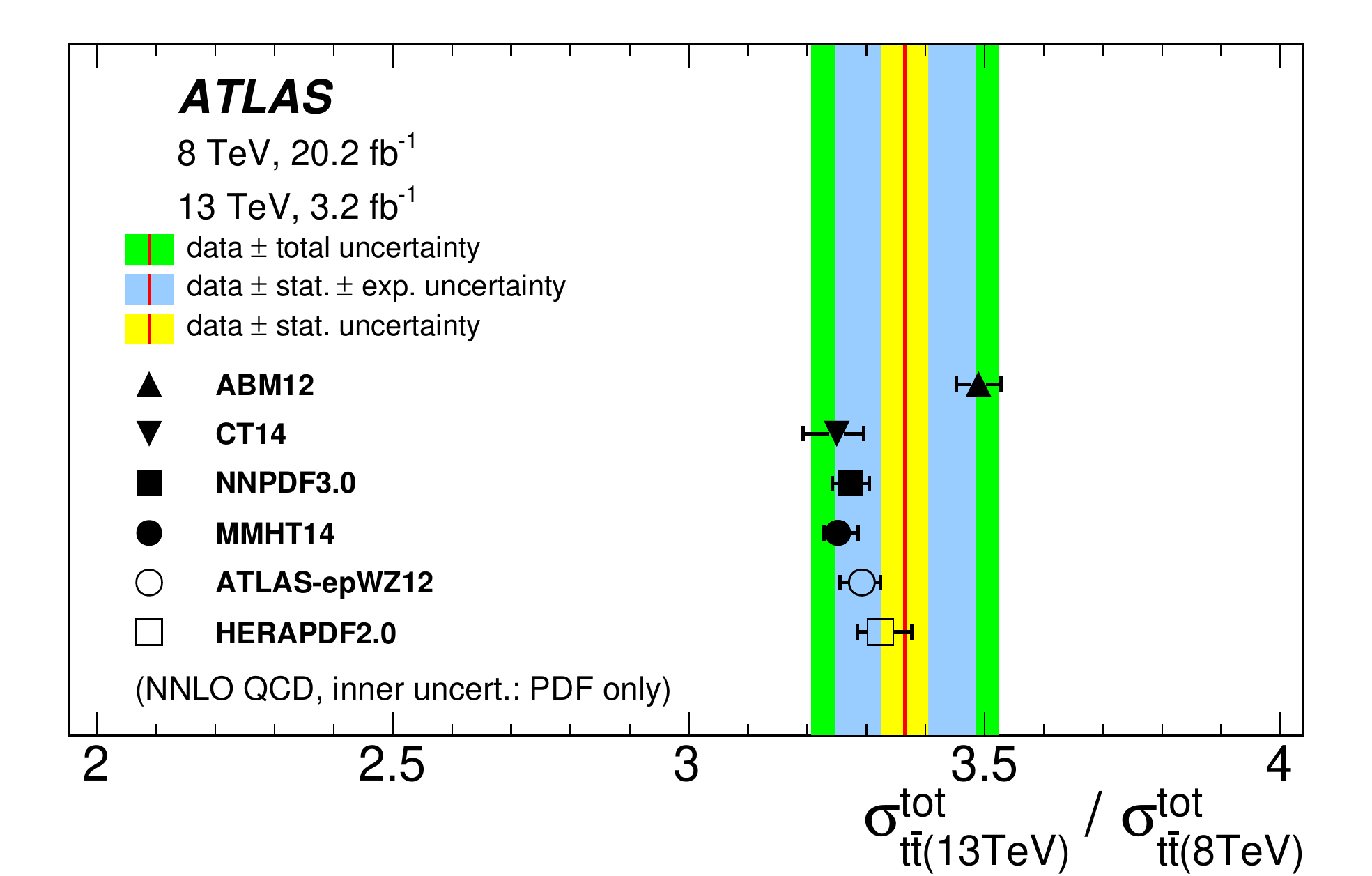}
  \hfill
  \includegraphics[width=0.5\textwidth]{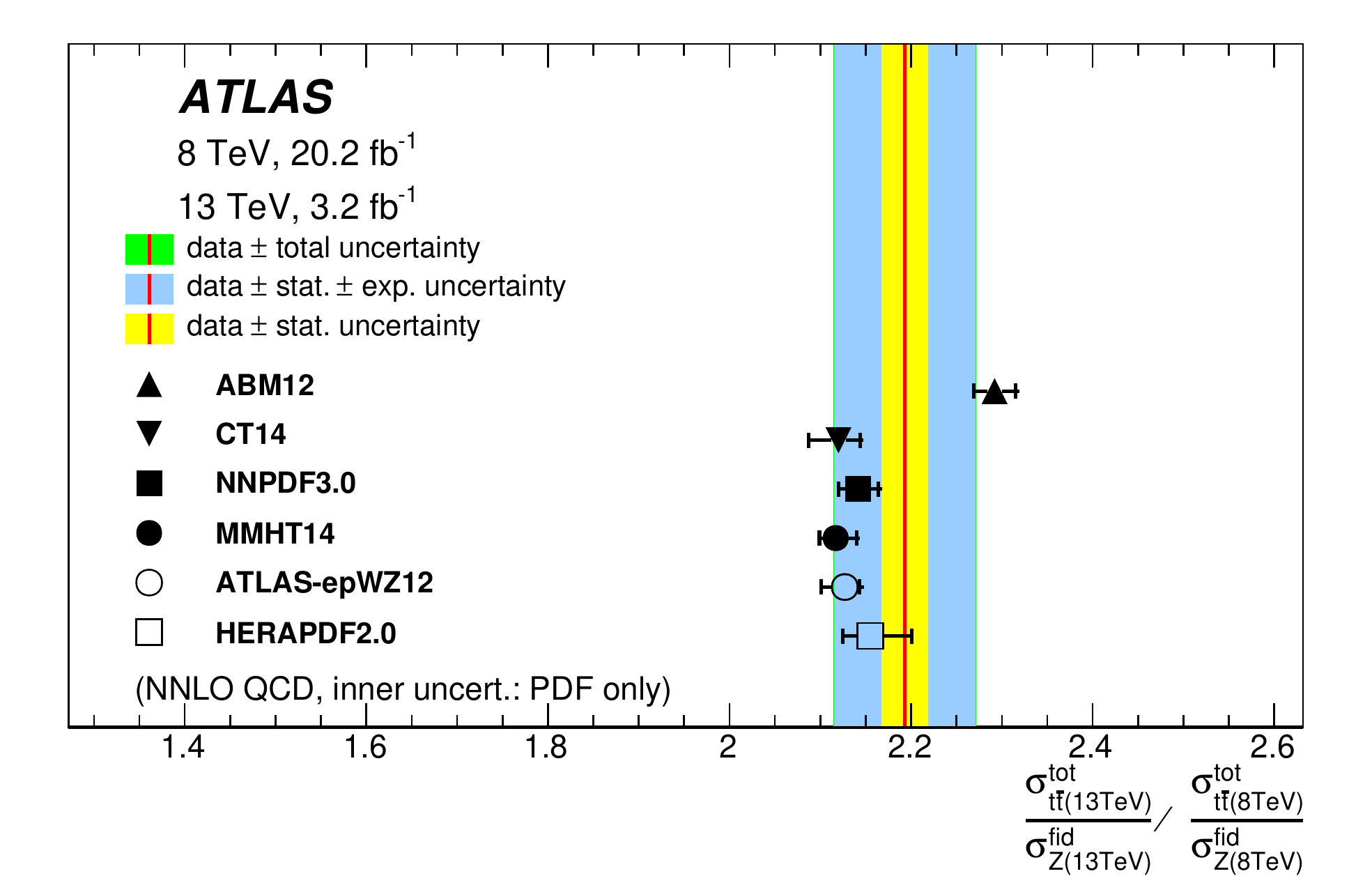}
  \caption{%
   Left: ratio of \ttbar cross sections at 8 and 13~TeV. Right: ratio of the ratios of \ttbar and $Z$-boson cross sections at 8 and 13~TeV. 
From Ref.~\cite{atlas-ttratios}.
  }
  \label{fig:atlas_ratios}
\end{figure}

Production rates of \ttbar pairs are powerful handles to constrain the parton distribution functions (PDF). This motivates for example the inclusive measurement at 5.02~TeV~\cite{tt5tev}, exploiting the fact that at relatively low CM energy the production of two heavy particles can only originate from incoming partons at the highest-$x$ tips of their spectra, where the gluon PDF is poorly constrained. 
I mentioned above that taking ratios can be an effective way to do more with the available data; an example is a clever study performed for the purpose of PDF extraction~\cite{atlas-ttratios}, where the ratios between \ttbar cross sections at different CM energies (Fig.~\ref{fig:atlas_ratios}, left), between \ttbar and $Z$-boson cross sections at the same CM energy, and ratios of ratios (Fig.~\ref{fig:atlas_ratios}, right) allow to get rid of the nuisances coming from luminosity uncertainty and other systematics.

\begin{figure}[!h!tbp]
  \includegraphics[width=0.47\textwidth]{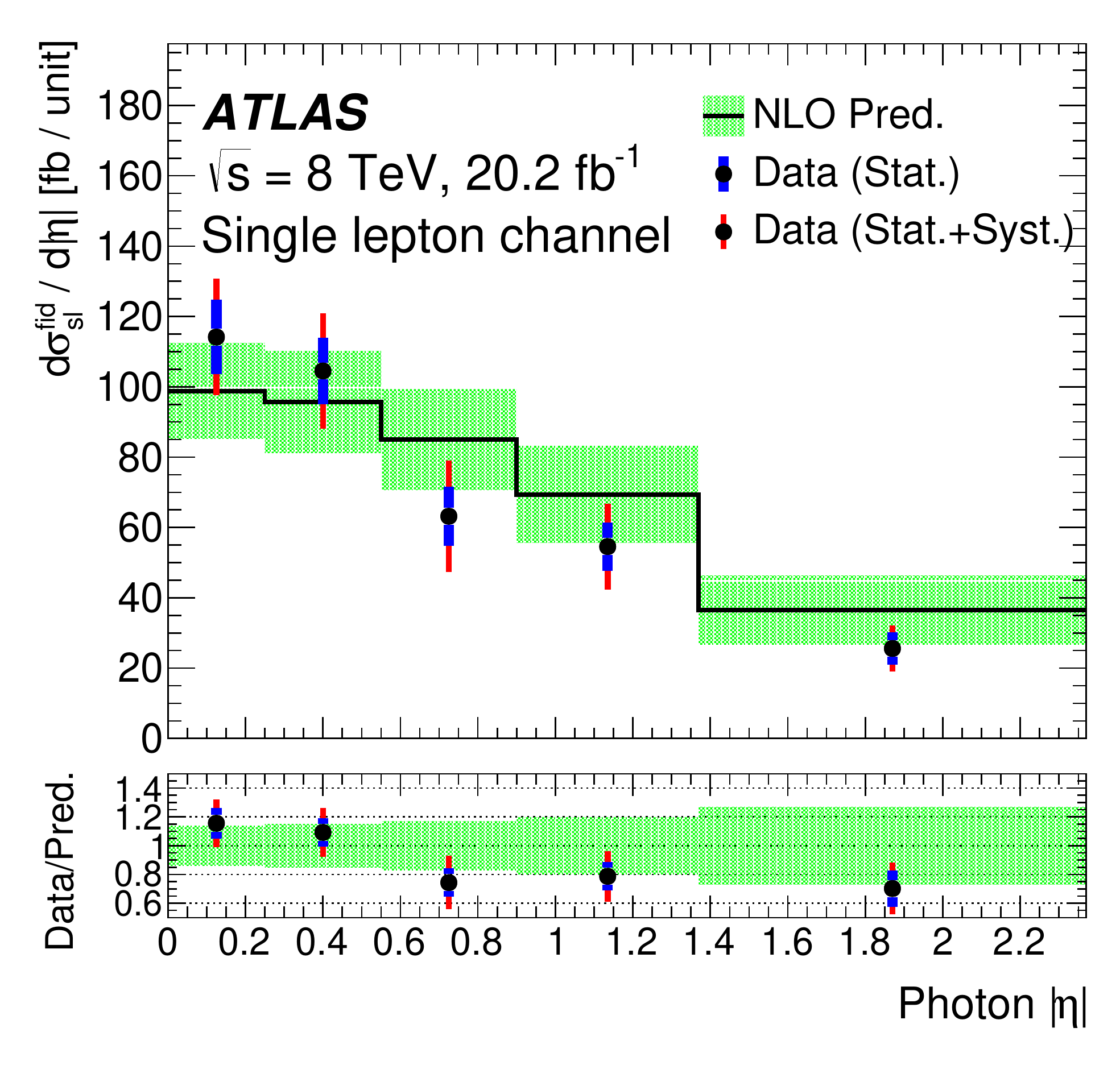}
  \hfill
  \includegraphics[width=0.5\textwidth]{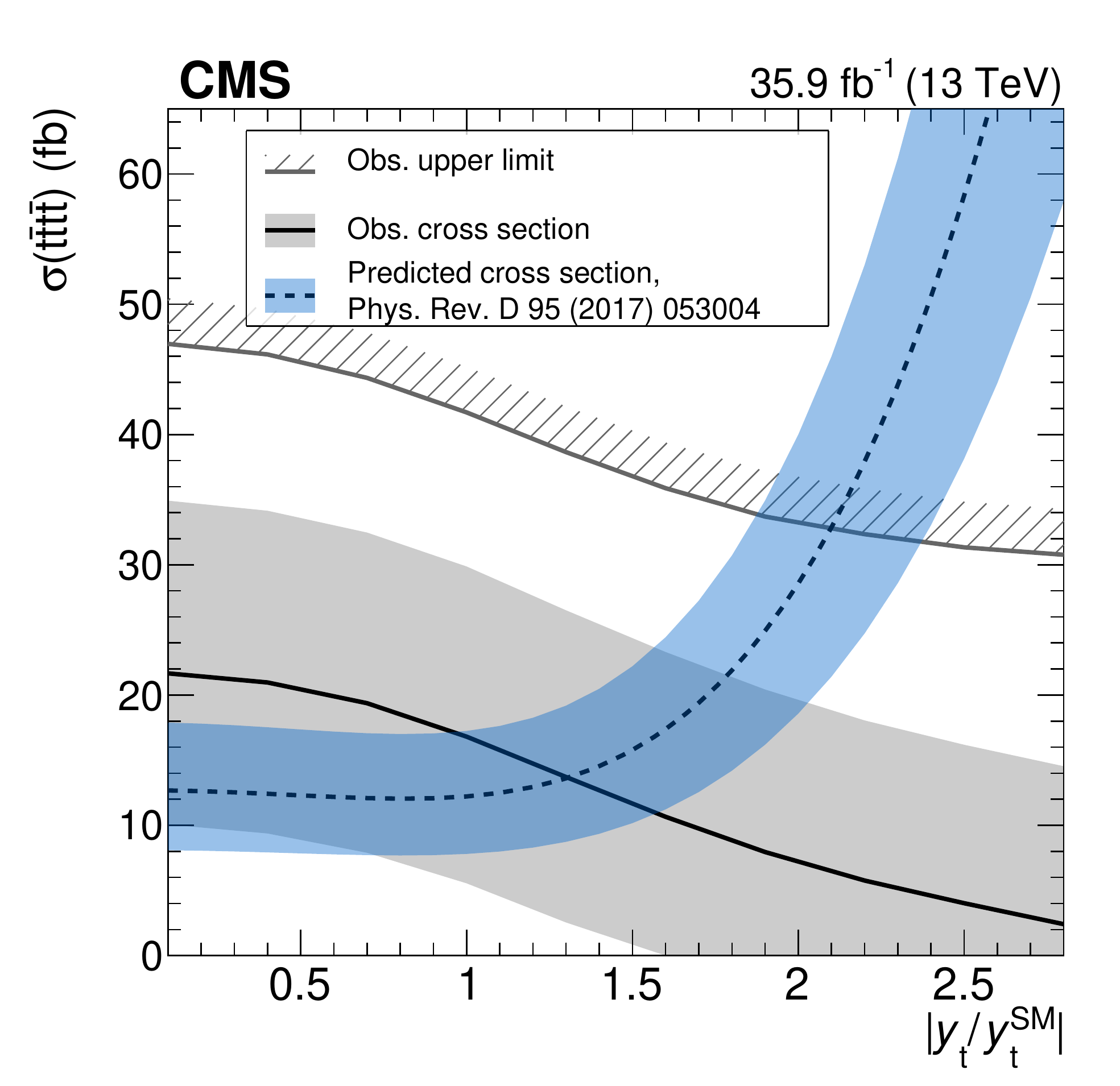}
  \caption{%
Left: differential cross section of $\ttbar\gamma$ production versus the pseudorapidity of the photon, from Ref.~\cite{atlas-ttg}.
Right: measurement in the plane formed by the cross section of $\ttbar\ttbar$ production and the Yukawa coupling of the top quark, from Ref.~\cite{cms-4t}.
  }
  \label{fig:ttg_4t}
\end{figure}

I mentioned differential measurements as another typical strategy to squeeze additional physics from the data, when statistics is abundant. In fact, differential cross-section meaurements are a booming business in the top-quark sector nowadays, with measurements performed at all CM energies, using most decay channels, in a great variety of observables at both parton and particle level~\cite{proc-ttbar-diff}. 
Constraining PDF (again), as well as Effective Field Theories (EFT), are among the main motivations for going differential in top-quark cross sections. 
A measurement of particular interest for EFT makes use of a process that used to be statistics-limited until recently: \ttbar in association with a hard photon ($\ttbar\gamma$)~\cite{atlas-ttg}.
Figure~\ref{fig:ttg_4t}, left, shows the differential cross section measured as function of the pseudorapidity of the photon, an observable that together with the transverse momentum (also measured in the same analysis) can be affected by modifications of the top-photon coupling. 

In general, several processes of the kind ``top pairs plus associated particles'' are interesting in EFT perspective and used to be considered rare until very recently. The cross sections of the $\ttbar V$ ($V=W,Z$) processes at 13~TeV, with the 2016 data, just passed the tipping point where statistical error is smaller than systematic~\cite{cms-ttv}. 
Another example is four-top production ($\ttbar\ttbar$), a rare process with large sensitivity to a variety of New Physics effects including anomalous top-Higgs couplings, where we are nearing sensitivity to the SM rate expectation~\cite{cms-4t} (Fig.~\ref{fig:ttg_4t}, right.)

As data accumulate, uncertainties on signal modeling becomes more and more of a hindrance for further progress in the top-quark sector. 
Here comes the ``Stop and Think'' attitude to the rescue. 
Turning the issue over its head, \ttbar measurements are more and more used to constrain the Monte Carlo (MC)  models~\cite{mc-tuning}, by using the aforementioned differential cross sections as input for MC tuning or performing dedicated ancillary measurements, such as the studies of colour flow in \ttbar events~\cite{colour-flow} and the measurement of the rates of additional heavy quarks~\cite{bachacou}.

\subsection{The single-top family, larger than ever}

\begin{figure}[!h!tbp]
  \center
  \includegraphics[width=0.8\textwidth]{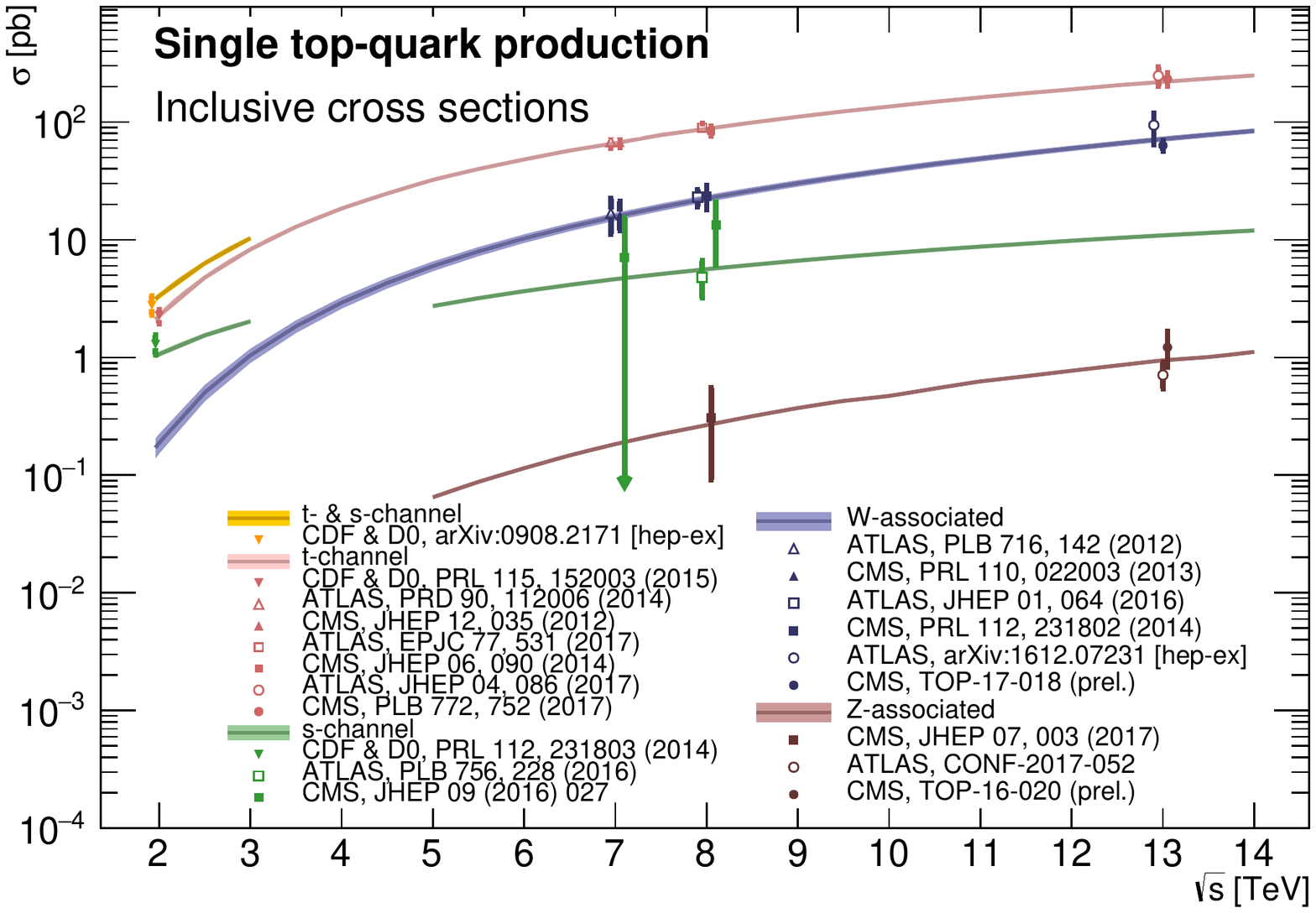}
  \caption{%
   Measurements of inclusive single top-quark cross sections in \pp and \ppbar collisions. 
From Ref.~\cite{giammanco}, where explanations and references are provided for the theory curves kindly provided by N. Kidonakis and J. Andrea.
  }
  \label{fig:stxsec}
\end{figure}

I estimate that the vast majority of the papers and talks about single-top physics produced before TOP2017 contain some variant of the sentence ``There are three production mechanisms for single top quarks in the Standard Model''~\footnote{I sinned too~\cite{cms-first-st}.}. 
But not anymore~\cite{proc-singletop}, thanks to the recent arrival of a little brother in the single-top family: the $Z$-associated production, or $tZ$, or $tZq$, measured by ATLAS at 13~TeV~\cite{atlas-zt-13tev} and CMS at 8 and 13~TeV~\cite{cms-zt} using the clean three-lepton final state where one lepton is assumed to come from a top quark decay and the other two are selected such to be compatible with a $Z$-boson decay. The ATLAS and CMS analyses at 13~TeV, based on the full 2016 dataset, both pass the conventional $3\sigma$ threshold on the incompatibility with the background-only hypothesis that allows to talk about ``evidence'' for the existence of the process~\footnote{The author is aware of the shortcomings of this convention, especially in cases where the signal expectation is precisely determined in the SM; see discussion in~\cite{dorigo}.}.

Also $W$-associated production (or $tW$) at 13 TeV has been in the spotlight at this conference, with a very precise inclusive cross section measured by CMS~\cite{cms-tw}, and the first ever differential cross sections measured by ATLAS~\cite{atlas-tw-diff}.

Figure~\ref{fig:stxsec} summarises the inclusive single top-quark cross section measurements at the time of the conference. 
The author notes in passing that an indication of the interest raised by the single-top research programme in the larger framework of top-quark physics may be found in the striking outcome of the TOP2017 Poster Prize. Out of 21 posters, only five were devoted to single top-quark physics but all three prize winners come from their ranks~\cite{poster-prizes}. 
It is left as an exercise to the reader to calculate the p-value of this occurrence, including the fact that their ranking in the Poster Prize corresponds to the hierarchy in cross section of the processes described.

\subsection{A new tool for Nuclear Physics}

The gluon PDF is known to be modified when the nucleons are bound in nuclei, with respect to free protons, and measurements of top-quark production in proton-nucleus collisions provide a powerful novel handle to constrain the nuclear gluon density at high virtualities in the unexplored high-$x$ region~\cite{denterria}. In addition, these measurements may serve as a baseline for future studies of the top quark as probe of the quark-gluon plasma formed in nucleus-nucleus collisions~\cite{apolinario}.

\begin{figure}[!h!tbp]
  \includegraphics[width=0.44\textwidth]{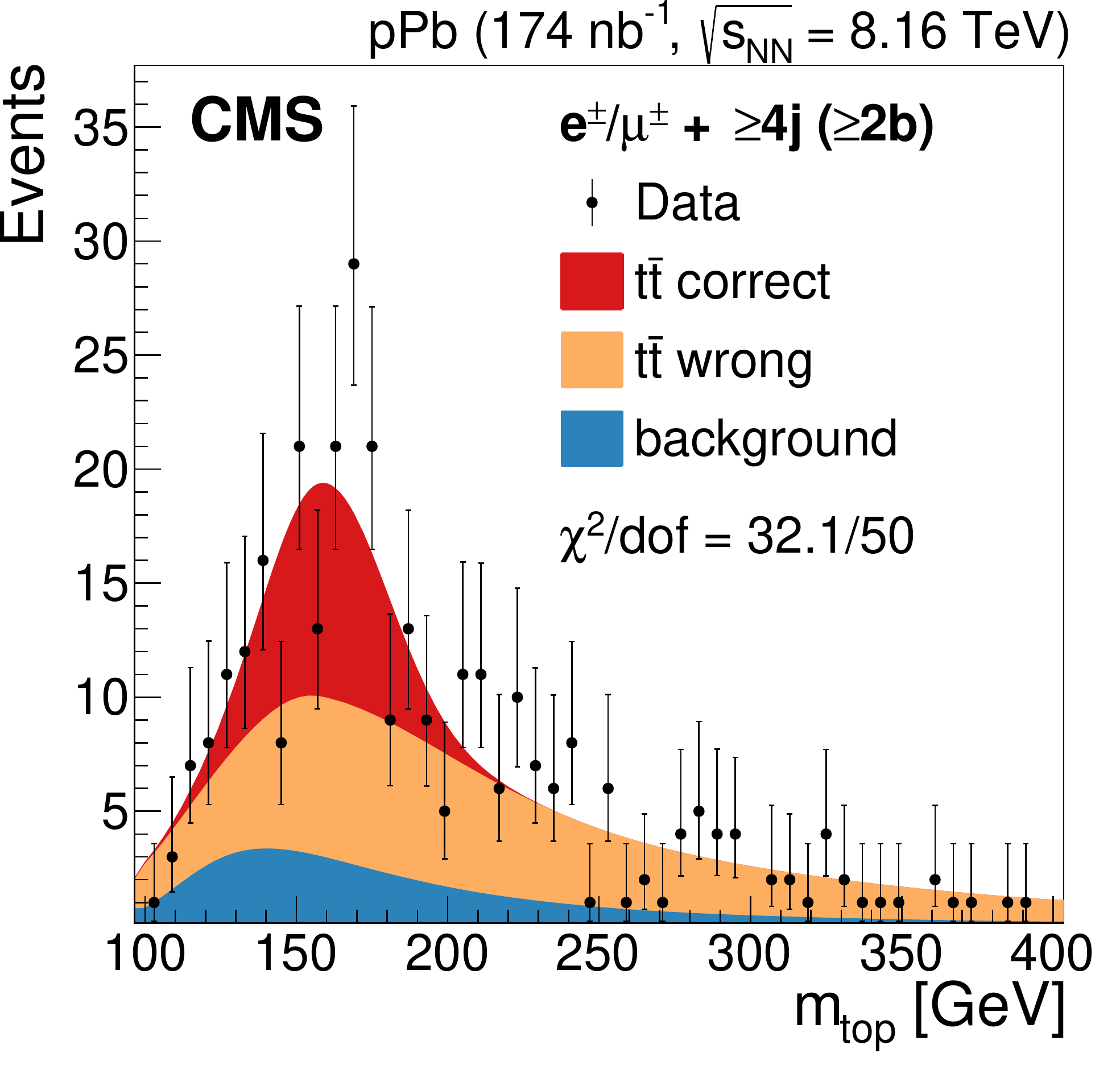}
  \hfill
  \includegraphics[width=0.55\textwidth]{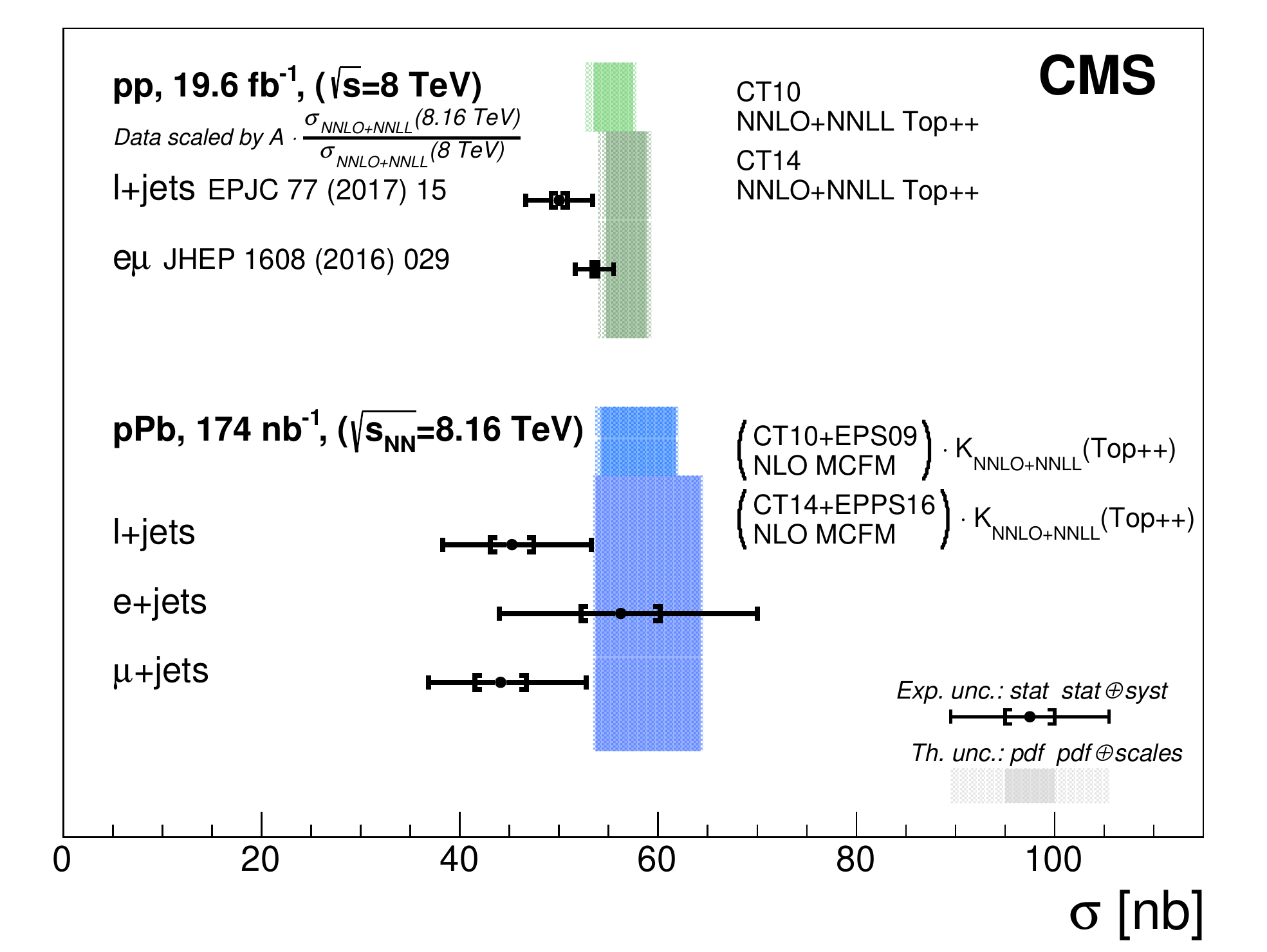}
  \caption{%
   Left: invariant mass distribution of the hadronic top-quark candidates in proton-lead collisions.
Right: CMS measurements of \ttbar cross sections at $\sqrt{s}=8$~TeV in proton-proton and $\sqrt{s_{NN}}=8.16$~TeV in proton-lead collisions.
From Ref.~\cite{ttpPb}.
  }
  \label{fig:pPb}
\end{figure}

This conference featured the first presentation of the first top-quark study ever appeared on the {\it nucl-ex} directory of arXiv~\cite{ttpPb}, based on 0.174~\pb of proton-lead collisions recorded in late 2016 (equivalent to 36~\pb of nucleon-nucleon collision data) at $\sqrt{s_{NN}}=8.16$~TeV. 
The analysis was performed in the single-lepton channel, with a fit to the invariant mass of the two untagged jets interpreted as $W$ boson decay products, using events with 0, 1 or 2 $b$-tagged jets. 
In this way, the background behaviour in this (so far unexplored) phase-space region and the $b$-tagging efficiency are evaluated {\it in situ} with only minimal assumptions, independent of prior inputs. As a validation, the outcome of the fit is used to model the signal and background invariant mass distributions of the hadronic top quark candidate (Fig.~\ref{fig:pPb}, left).
The excess of events with respect to the background-only hypothesis is quantified with a variety of methods and assumptions, yielding a significance of more than $5\sigma$ even under the most conservative assumptions. The measured top-pair cross section is consistent with the expectations from scaled \pp data as well as perturbative QCD predictions (Fig.~\ref{fig:pPb}, right).

\section{``Isn't the top quark just two numbers?''}
\label{sec:two-numbers}

During the keynote speech that opened the conference~\cite{craig}, we were told about a prominent General Relativist who dismisses the entire top-quark research programme with the question that titles this section. 
The audience, being a TOP conference, was at the same time amused and horrified by this anecdote. 
As another speaker remarked~\cite{melnikov}, the Particle Data Group seems to consider that a lot more than just two numbers are worth being reported for our dearest quark~\cite{pdg}.

But let's take this question seriously, in this section, for sake of argument. 
The two numbers that the unnamed colleague was referring to, are the mass and the decay width of the top quark. 
I argue that the values of those two numbers are precisely what makes the top quark so special to mobilise a significant fraction of the collider community and motivate a yearly conference with more than one hundred attendees!

\subsection{First number: mass}

Let's start from the mass. 
It is so large that only two colliders were able to study the top quark so far. So large that in comparison all other fermions are almost massless. 
According to how the SM describes the origin of fermion masses (which is still valid in many of its extensions), we can relate the top-quark mass to a Yukawa coupling:
\begin{equation}
y_t = \sqrt{2}\frac{m_t}{v},
\label{eq:mass}
\end{equation}
where $v=246.2196(1)$~GeV~\cite{pdg} is the vacuum expectation value of the Higgs field. Plugging the direct measurements~\cite{world_average,atlas_mass,cms_mass,tevatron_mass} of $m_t$ in this equation, we get a value so strikingly close to 1 that many take it as a hint that the top quark may play some special role in the electroweak symmetry breaking. 

\begin{figure}[!h!tbp]
  \includegraphics[width=0.5\textwidth]{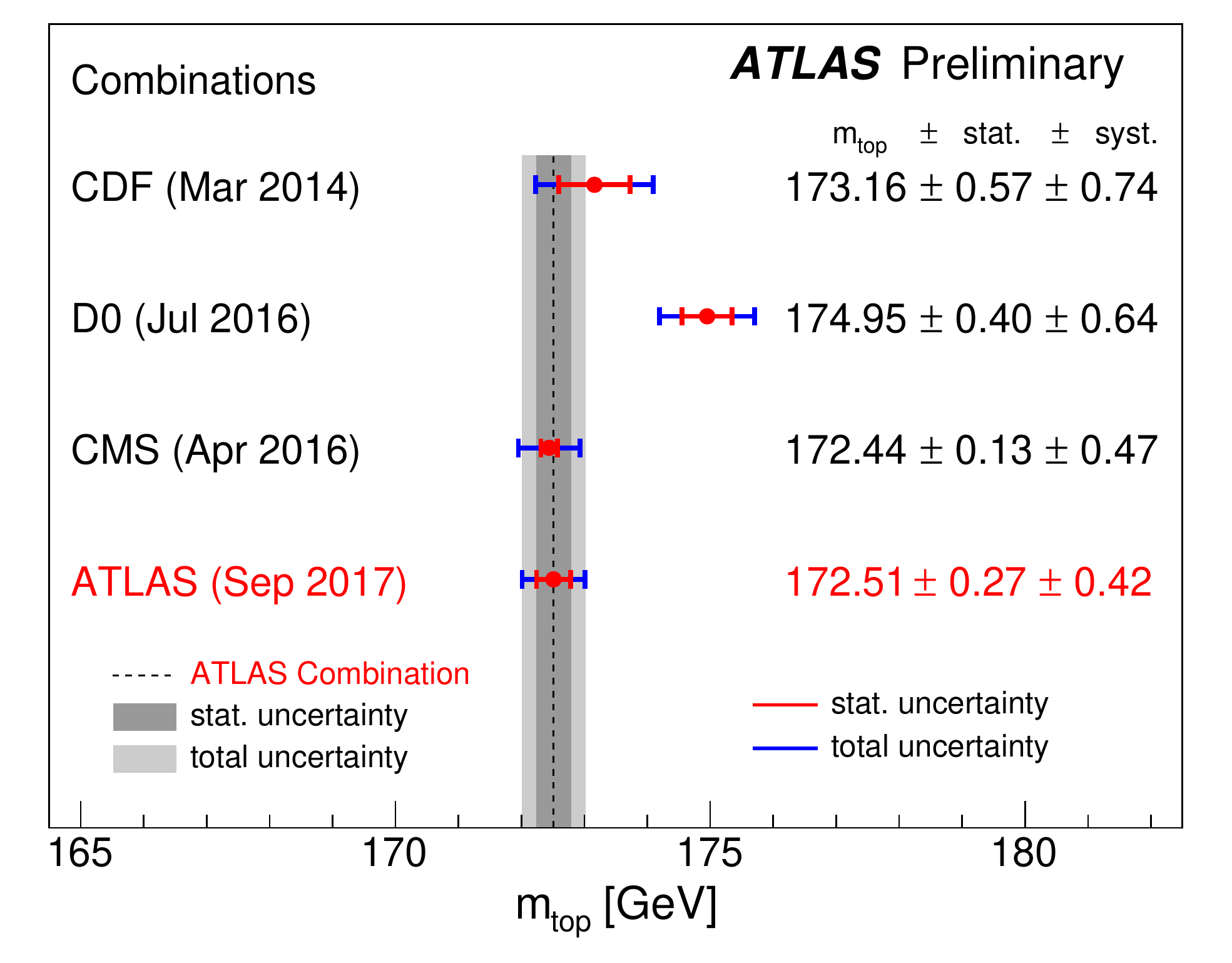}
  \hfill
  \includegraphics[width=0.45\textwidth]{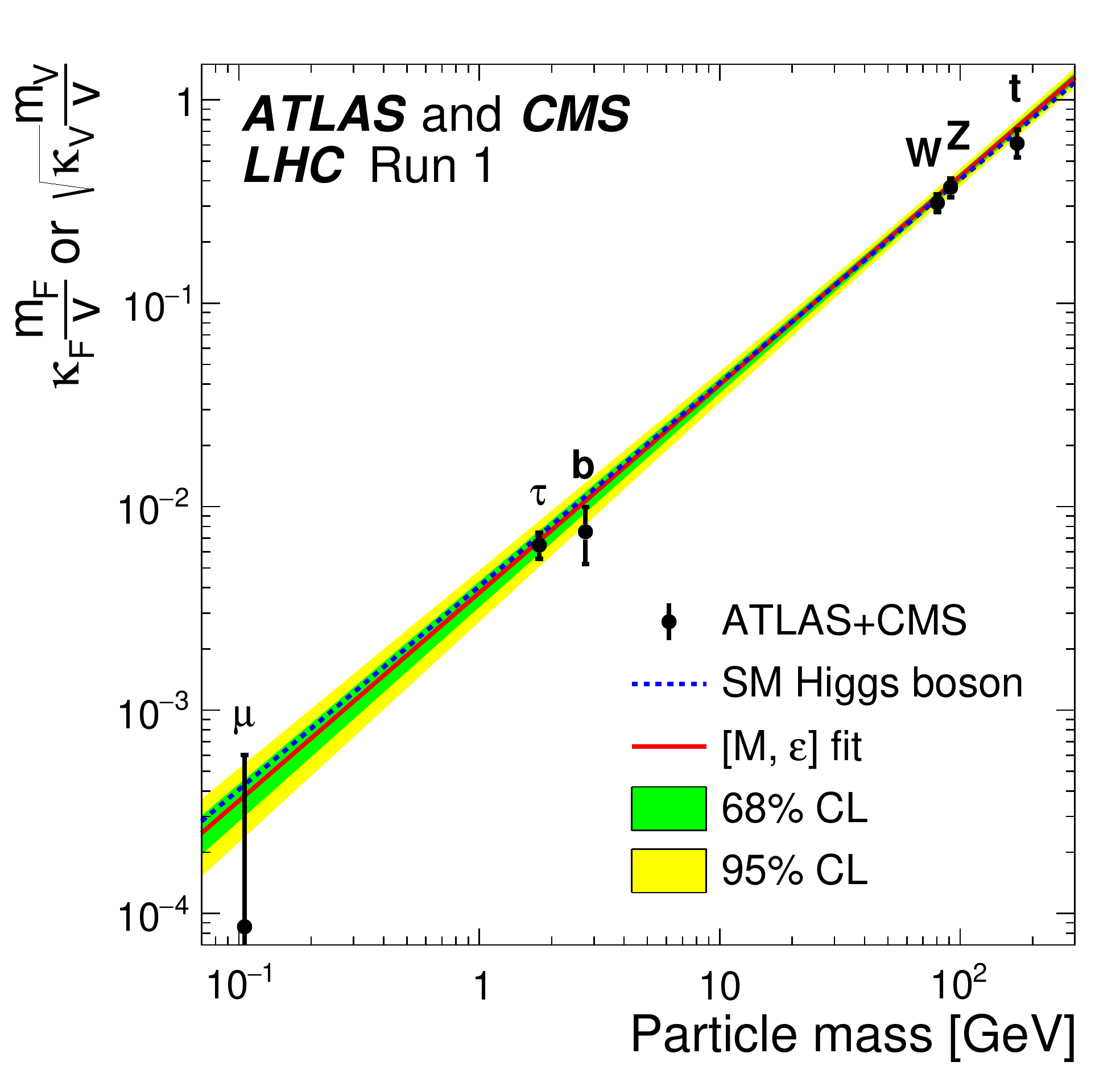}
  \caption{%
Left: individual combinations of the top-quark mass measurements by the Tevatron and LHC collaborations~\cite{atlas_mass,cms_mass,tevatron_mass}, as summarised in Ref.~\cite{atlas_mass}. 
Right: Higgs coupling strengths versus particle masses, from LHC Run-1 data~\cite{higgs-combo-run1}.
  }
  \label{fig:mass}
\end{figure}

So, definitely, this first number is very worth being studied, at the best possible precision and from multiple angles. And in fact that's what a significant subset of this community is doing~\footnote{The reader will find elsewhere~\cite{pole_mass,melnikov} a summary of the interesting and very divisive ``pole mass versus MC mass'' issue (or non-issue?) that this margin is too narrow to contain.}. 
Things go so fast in this area that, just three years after its release, the first and so far only world combination of $m_t$~\cite{world_average} is already surpassed in precision by the individual combinations of ATLAS, CMS, and Tevatron data~\cite{atlas_mass,cms_mass,tevatron_mass}. 
In the occasion of this conference, ATLAS released its legacy Run-1 measurement of $m_t$~\cite{atlas_mass}, in the most precise channel (single-leptonic) and with the 8~TeV dataset, expected to finally trigger a new round of world combination. The same studies needed for the combination will hopefully clarify whether the deviation between D0 and the LHC experiments (see Fig.~\ref{fig:mass}, left) is a statistical fluke or a hint of some unaccounted systematic effect. In the same occasion CMS released the first measurement based on the 2016 dataset at 13~TeV~\cite{cms_mass}.

Measuring both sides of Eq.~(\ref{eq:mass}) is a crucial test of the SM. The right-hand side is now known at the level of 0.3\%, and the incompatibility of the left-hand side with zero is currently established at the level of ``evidence'', i.e. more than $3\sigma$, by a variety of individual inputs. Indirectly, a more than $4\sigma$ evidence for the top-Higgs coupling is obtained trough a global fit to the full Run-1 data (ATLAS+CMS) in a variety of decay channels and associated objects~\cite{higgs-combo-run1} (Fig.~\ref{fig:mass}, right), where the sensitivity to $y_t$ mostly originates from the loops in the $gg\to H$ and $H\to\gamma\gamma$ vertices (signal regions enriched in \ttH events are also included in the global fit, but are less impactful). 
But this interpretation of the data relies critically on some assumptions about what is allowed to run in the loops, which strongly motivates the hunt for the \ttH process as its cross section can be directly translated into a $|y_t|$ measurement~\footnote{When ignoring the tiny higher-order electroweak corrections, to which current measurements are insensitive, the cross section of \ttH production only gives sensitivity to the modulus of the coupling.}. The current Run-2 status shows independent direct evidence for this process from both ATLAS and CMS~\cite{tth-evidences}.

\subsection{Second number: width}

Also the width of the top quark is a remarkable number: if the SM holds, from its mass one can deduce a width of more than one~GeV~\cite{pdg}, corresponding to a lifetime of $\approx 0.5\times 10^{-24}$~s, unusually short for the weak decay of a fermion. This is due to the fact that $m_t$ is larger than the sum of the $W$ and $b$ masses, therefore there is no barrier to overcome and we have a two-body decay $t\to Wb$ with a real $W$ boson, instead of the usual three-body decay mediated by a virtual $W$ boson. 

Tevatron and LHC experiments performed several measurements directly or indirectly related to the top-quark width~\cite{proc-properties}. 
For this conference, ATLAS released a direct measurement in the single-lepton channel using the 8~TeV dataset~\cite{atlas-width}. 
Two observables are used: the invariant mass of the lepton and the $b$ jet from the leptonically-decaying top quark, and the distance between the other $b$ jet and the closest light jet from the hadronically-decaying $W$ boson. The result agrees with the SM within the 50\% experimental uncertainty.

\begin{figure}[!h!tbp]
  \center
  \includegraphics[width=0.44\textwidth]{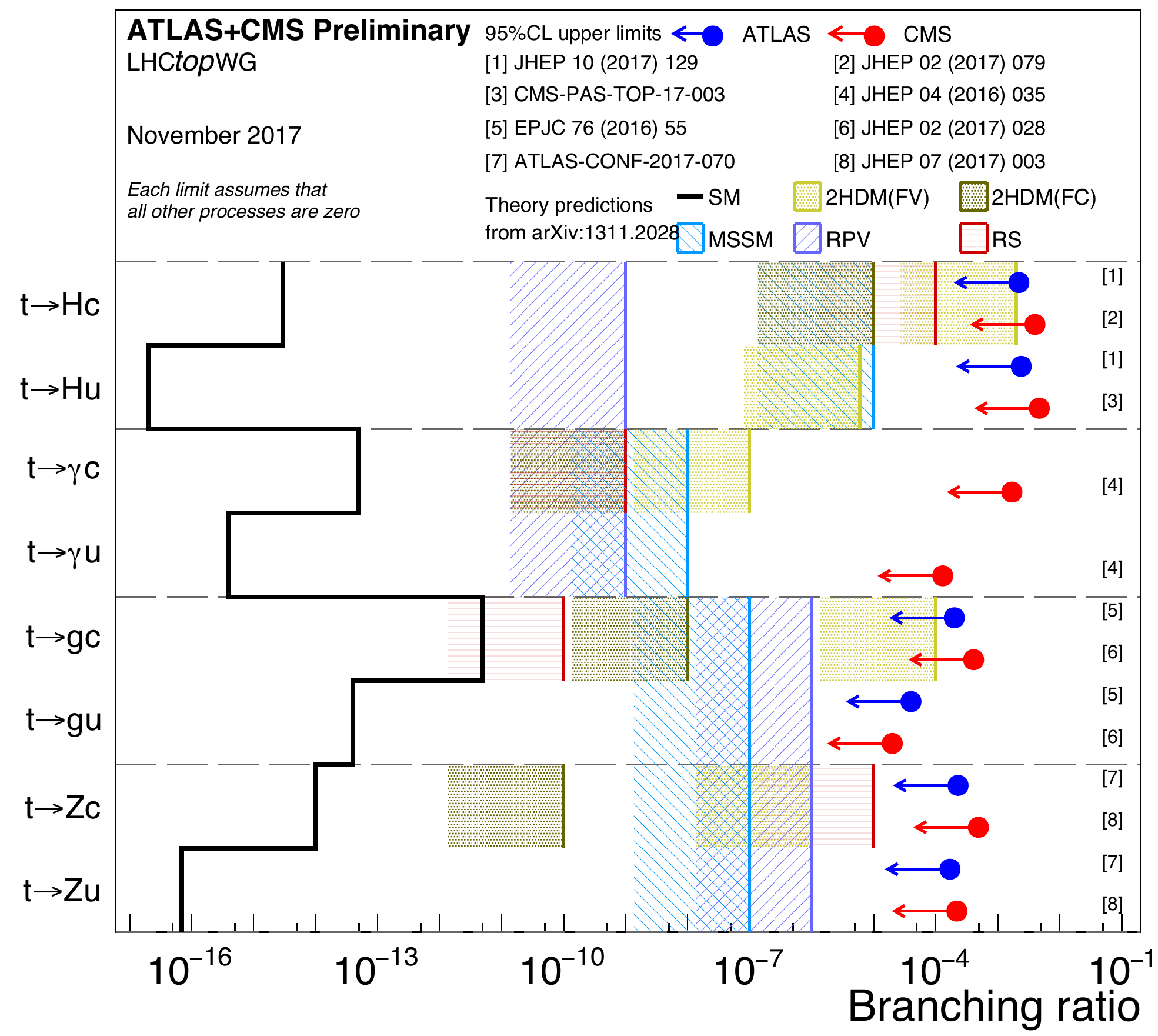}
  \includegraphics[width=0.44\textwidth]{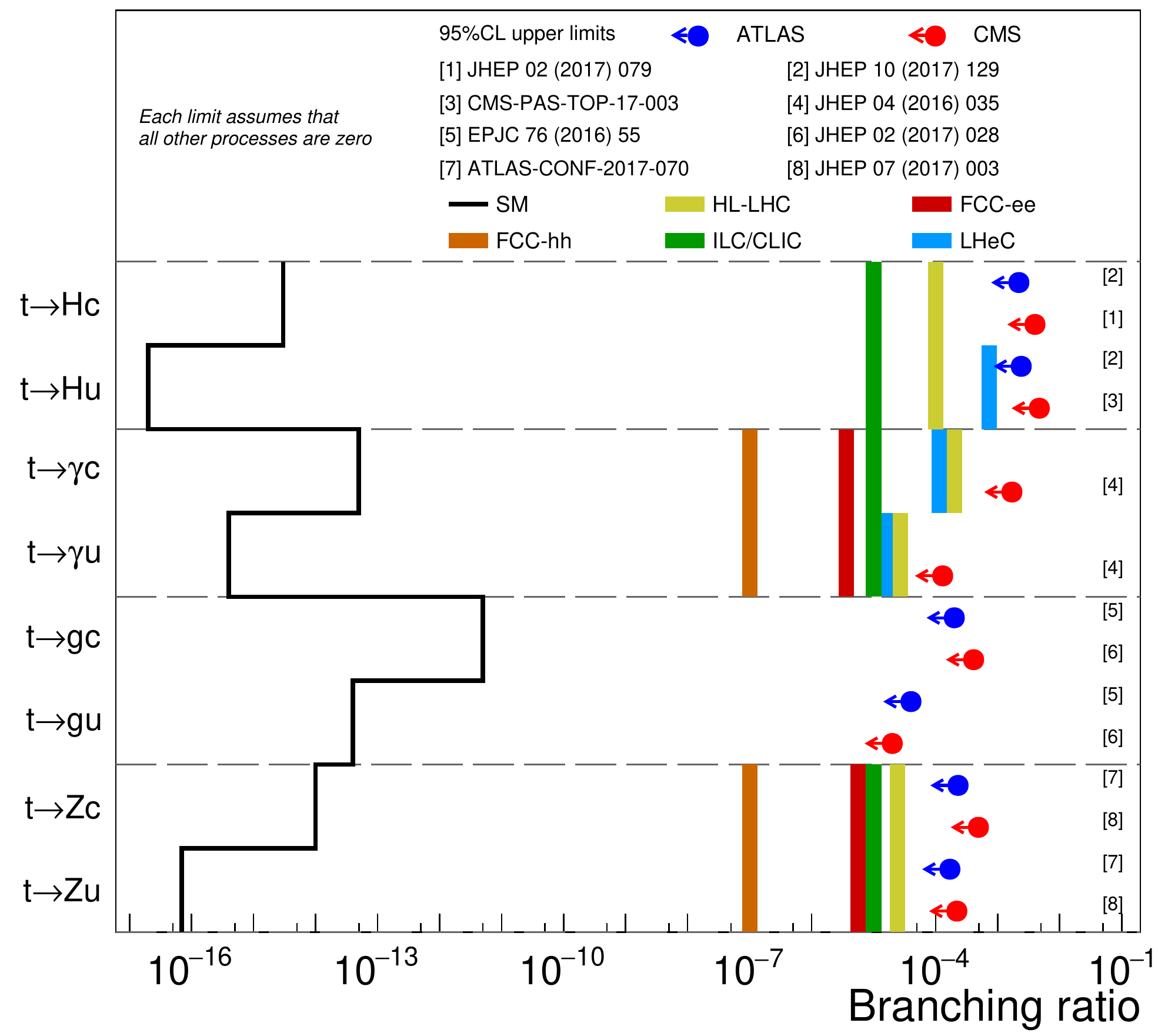}
  \caption{%
   Current limits on FCNC couplings of the top quark, expressed in terms of equivalent branching ratio.
Left: compared to several models, from the LHC Top Working Group~\cite{lhctopwg}. Right: expectations for future accelerators, from Ref.~\cite{skovpen}.
  }
  \label{fig:fcnc}
\end{figure}

Contributions to the width may come from new decay channels mediated by Flavour Changing Neutral Currents (FCNC), and this conference saw a burst of fresh constraints based on 2016 data~\cite{skovpen}, in the $t\to Zq$~\cite{fcnc-tzq} and $t\to Hq$~\cite{fcnc-thq} channels ($q=u,c$). 
Figure~\ref{fig:fcnc} summarises the current limits on these FCNC processes.

\section{Angular distributions and asymmetries}
\label{sec:asymmetries}

A consequence of the large width of the top quark is that its lifetime is shorter than both the spin-decoherence and hadronisation timescales, which means that most of its polarisation is passed to its decay products. Most notably, the experimentally clean charged leptons~\footnote{Common lore says that hadronisation washes away any such information, hence $b$ quarks are usually disregarded as a possible probe for spin-related information. However, this statement is not true for the $O(10\%)$ fraction of $b$ quarks that hadronise into baryons. Reference~\cite{lambda_b} suggests how to use this fact to characterise possible new resonances decaying preferentially into heavy quarks, and how to calibrate the method using clean \ttbar samples.} carry a lot of powerful information through their angular distributions, correlations with other final-state objects, and angular asymmetries. This is exploited by analyses of \ttbar spin correlations, top-quark polarisation in \ttbar (where it is expected to be 0) and single top (expected to be $\approx 1$), and $W$-boson helicities. In common between these heterogeneous analyses, reviewed elsewhere~\cite{proc-properties,proc-properties-st}, is that the observables of interest are angular distributions or angular asymmetries.

\begin{figure}[!h!tbp]
  \includegraphics[width=0.43\textwidth]{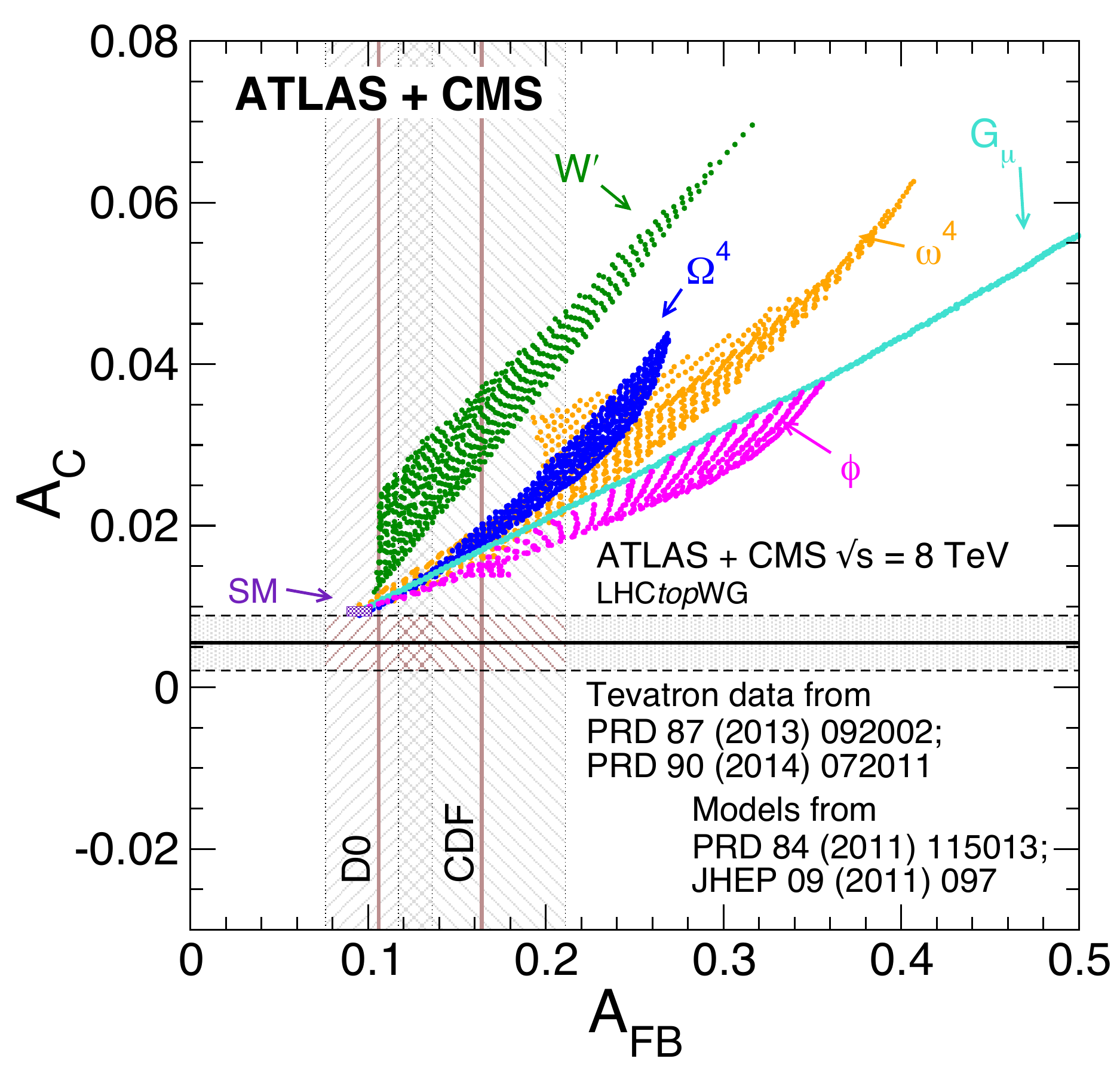}
  \hfill
  \includegraphics[width=0.56\textwidth]{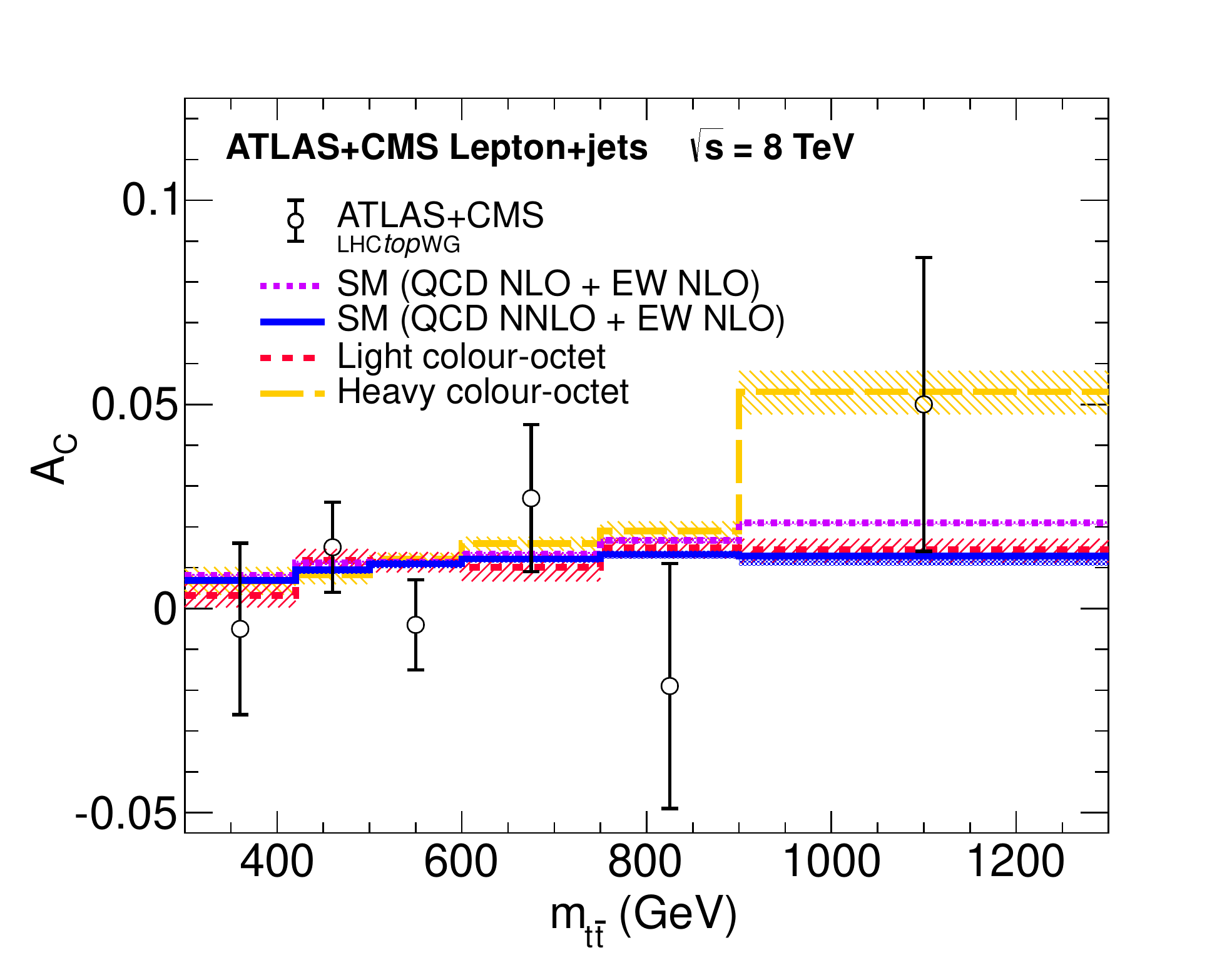}
  \caption{%
   Left: inclusive $A_C$ versus inclusive $A_{FB}$ in \ttbar events at LHC and Tevatron, respectively. 
Right: differential $A_C$ as function of \ttbar invariant mass.
From Ref.~\cite{lhc-ac}.
  }
  \label{fig:Ac}
\end{figure}

Talking about a different kind of asymmetries, most of the TOP editions so far have been dominated by discussions of the forward-backward asymmetry, $A_{FB}$, of \ttbar events in \ppbar collisions, which for some time has been in tension with the expectations. 
Due to the symmetry of its beams, LHC can cross-check the Tevatron measurements of $A_{FB}$ only by measuring the related central-edge asymmetry (commonly called charge asymmetry, for some reason), $A_C$. Both quantities have been measured inclusively and differentially~\cite{proc-properties}. Although the experimental measurements of $A_{FB}$ are now fairly compatible with the latest expectations, the bygone excesses had the beneficial long-term effect of attracting the attention of the community to the sensitivity of these observables to new physics. 
This motivated the ATLAS and CMS analysts to join forces for the first joint paper on $A_C$ measurements~\cite{lhc-ac}, featuring inclusive measurements at 7 and 8~TeV and a differential measurement as a function of the \ttbar invariant mass at 8~TeV (Fig.~\ref{fig:Ac}). For the reader interested in the sociology of HEP, it can be remarked that never so far the ATLAS and CMS collaborations had considered a top-quark result worth the editorial burden of a joint peer-reviewed publication (although we had several joint notes already~\cite{lhctopwg}.)

\section{Conclusions}
\label{sec:conclusions}

The LHC collaborations released a swarm of new results at the TOP2017 conference: legacy Run-1 precision measurements that exploit the ultimate calibrations; measurements and searches at 13~TeV using the large 2016 dataset; and the exploitation of some ``special runs'' that turned out to be fit for top physics too, like the short ``reference pp run'' at $\sqrt{s}=5.02$~TeV in 2015~\footnote{Superseded by the ten times larger dataset at the same CM energy collected in November 2017.} and the pPb run at $\sqrt{s_{NN}}=8.16$~TeV in 2016. 
And in spite of a very reduced work force, six years after the end of operations, Tevatron experiments continue to contribute new results in this area. 

The experimental programme of the conference included reports on the performances of the ATLAS and CMS experiments~\cite{lhc-perf}. It is interesting to note that pileup, one of the highest worries for the current and future LHC physics programme, was never reported as a limiting factor in any of the LHC talks at this conference, thanks to the work successfully invested in mitigation strategies. 
We also had reports from outside of top quark physics proper, to put our research area in a broader context: updates about SUSY and Dark Matter searches~\cite{guest} and searches for resonances and vector-like new quarks~\cite{gerrit}. 
I conclude by mentioning the intense discussions on the subject of algorithms for boosted top quarks~\cite{boosted}, some of them borrowing novel ideas from outside of HEP. These new developments allow to tap more statistics in the extremely important phase-space corner sampled by the highest-energy top quarks, and will thus become even more important at future accelerators.

\Acknowledgements
I am grateful to the International Advisory Committee of the TOP2017 conference for the kind invitation to give this summary, 
and to all the speakers and attendees for making it a very pleasant experience.

\end{document}

%% file: econfmacros.tex
\def\beq{\begin{equation}}
\def\eeq#1{\label{#1}\end{equation}}
\def\eeqn{\end{equation}}

\def\beqa{\begin{eqnarray}}
\def\eeqa#1{\label{#1}\end{eqnarray}}
\def\eeqan{\end{eqnarray}}

\let\bar=\overbar

\def\Dslash{\not{\hbox{\kern-4pt $D$}}}
\def\dslash{\not{\hbox{\kern-2pt $\del$}}}

\def\msb{{\bar{\ssstyle M \kern -1pt S}}}

